# Effect of Noise and Modeling Errors on the Reliability of Fully 3D Monte Carlo Reconstruction in SPECT

D. Lazaro, Z. El Bitar, V. Breton and I. Buvat

*Abstract*— We recently demonstrated the value of reconstructing SPECT data with fully 3D Monte Carlo reconstruction (F3DMC), in terms of spatial resolution and quantification. This was shown on a small cubic phantom (64 projections 10 x 10) in some idealistic configurations. The goals of the present study were to assess the effect of noise and modeling errors on the reliability of F3DMC, to propose and evaluate strategies for reducing the noise in the projector, and to demonstrate the feasibility of F3DMC for a dataset with realistic dimensions. A small cubic phantom and a realistic Jaszczak phantom dataset were considered. Projections and projectors for both phantoms were calculated using the Monte Carlo simulation code GATE. Projectors with different statistics were considered and two methods for reducing noise in the projector were investigated: one based on principal component analysis (PCA) and the other consisting in setting small probability values to zero. Energy and spatial shifts in projection sampling with respect to projector sampling were also introduced to test F3DMC in realistic conditions. Experiments with the cubic phantom showed the importance of using simulations with high statistics for calculating the projector, and the value of filtering the projector using a PCA approach. F3DMC was shown to be robust with respect to energy shift and small spatial sampling offset between the projector and the projections. Images of the Jaszczak phantom were successfully reconstructed and also showed promising results in terms of spatial resolution recovery and quantitative accuracy in small structures. It is concluded that the promising results of F3DMC hold on realistic data sets.

## I. INTRODUCTION

IN Single Photon Emission Computed Tomography (SPECT), the use of iterative reconstruction algorithms is appealing to account for the 3D physical effects involved in the imaging process, namely scatter and detector response. The projector **R** involved in iterative reconstruction can be modeled either using approximate analytical models [1,2] or using Monte Carlo simulations [3]. We recently revisited the Monte Carlo approach and demonstrated, on a small cubic phantom, that fully 3D Monte Carlo (F3DMC) reconstruction using a 3D projector estimated by high statistics Monte Carlo simulations was worthwhile to improve spatial resolution, signal-to-noise ratio and recovery of local tracer concentration [4,5]. In the present study, we assessed the effect of noise and modeling errors on the reliability of the reconstruction for a dataset of small dimensions hence easy to handle, and compared two different strategies for reducing noise in the projector. We also showed the feasibility of F3DMC on a realistic dataset using a simulation of a Jaszczak-like phantom.

## II. MATERIAL AND METHODS

### A. Monte Carlo simulations

Monte Carlo simulations were performed using the code GATE, which has been recently validated for various configurations in SPECT and PET (e.g., [6,7]). Photon transport within the collimator was simulated using the Monte Carlo approach, ensuring an accurate model of the imaging system sensitivity. No variance reduction techniques were used.

#### 1) Simulated configurations

Two phantoms were considered (Figure 1):
- a cubic phantom consisting of 3 Tc-99m line sources and 3 Tc-99m point sources embedded in air, water and bone (Fig. 1a). Different activity concentrations were set in the point sources and line sources while there was no activity in the background,
- a 10 cm diameter cylindrical phantom (Fig. 1b), filled with water, including 6 cylinders having diameters of 4.8, 6.4, 7.8, 9.5, 11.1 and 12.7 mm. The smallest five cylinders were filled with water whereas the biggest one contained bony material. The bony cylinder contained no activity. Tc-99m activity (2.08 MBq/ml) was set in the cylindrical phantom, while the 5 water filled cylinders contained 8.32 MBq/ml of Tc-99m.

The cubic and cylindrical phantom volumes were sampled on a 10×10×10 voxel grid (1 cm x 1 cm x 1 cm voxels) and on a 64×64×64 voxel grid (3.125 mm × 3.125 mm × 3.125 mm voxels), respectively. For the cubic phantom, 64 projections 10×10 (pixel size: 1 cm) were simulated (radius of rotation R=12 cm). For the cylinder, 64 projections 64×64 (pixel size: 3.125 mm) were simulated (R=12 cm). In both cases, the gamma camera characteristics mimicked those of an AXIS (Philips) equipped with an LEHR collimator. About 4.5 and 27.9 billion photons were simulated for the cubic phantom and the cylindrical phantom, respectively.

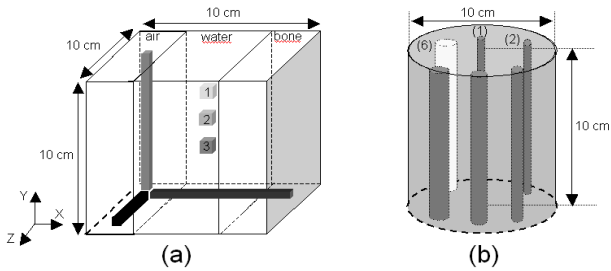

Fig. 1. (a) cubic phantom made of 3 attenuating media and in which Tc-99m line and point sources were inserted as shown; (b) water cylinder including 5 Tc-99m cylinders of different diameters and a bony cylinder.

*2) Projector calculation*

For each phantom, a uniform Tc-99m activity distribution within the phantom was simulated. 30 billion photons were simulated and 6.9 million photons were detected between 126 and 154 keV for the cubic phantom. These values corresponded to simulating 30 million counts per voxel and detecting an average of 1078 counts per pixel. The relative uncertainty $\Delta r_{ij}/r_{ij}$ associated with each projector element $r_{ij}$ was thus estimated as about 0.003 % (assuming that detected counts were uniformly distributed over the whole detection head). For the cylinder, about 64 billion photons were simulated (244140 counts per voxel) and about 14 million photons were detected between 126 and 154 keV (average of 53 per pixel). The corresponding relative uncertainty $\Delta r_{ij}/r_{ij}$ was thus about 0.25 %. In this latter case, the simulation took about 5 days on a cluster of 40 biprocessors Pentium III 1 GHz. For each phantom, the projector **R** was then deduced [4].

*B. Image reconstruction and assessment*

Data collected in the 126-154 keV energy window were reconstructed with F3DMC, implicitly correcting for scatter, attenuation and finite spatial resolution. Ordered Subset Expectation Maximisation (OSEM) was used as the iterative reconstruction algorithm involved in F3DMC (4 subsets, 14 iterations) [8].

To assess the accuracy of the reconstructed images, spatial resolution, signal-to-noise ratio (calculated from 20 replicates of noisy projections), activity ratios and absolute activity measurements were measured. For the cubic phantom, volumes of interest (VOI) consisting of the hottest 4 pixels in each Tc-99m line were considered to calculate activity ratios between the Z and Y lines, and between the Z and X lines. For the cylinder, activity ratios between each cylinder and the background activity were calculated, by considering cylindrical VOIs having diameters equal to the actual diameter of the different cylinders. A background VOI of 8 pixels in diameter was drawn at the centre of the cylinder. Activity ratios were also calculated on the original activity images used for the simulation, to separate the effect of sampling and the effect of reconstruction in the bias affecting the activity ratios. Absolute quantitation was assessed by calculating the percentage of recovered activity (%RA) for each phantom as follows: %RA = 100*(estimated phantom activity)/(simulated phantom activity), where the estimated phantom activity was calculated as the number of counts in the reconstructed images divided by the simulated acquisition time.

*C. Study of F3DMC robustness with respect to noise in the projector and modeling errors*

Using the cubic phantom, results obtained with projectors corresponding to 1 billion and 30 billion simulated counts were compared. Two strategies of filtering the projector to reduce the noise were also considered:

- principal component analysis (PCA) [9] of the projector. Assuming an acquisition of P projections of N × N pixels and an object of N × N × N voxels to be reconstructed, the projector of dimensions ($PN^2$, $N^3$) is first reorganized as P vectors of $N^5$ elements. A PCA analysis of these P vectors is performed, yielding P eigenvalues associated with P eigenvectors. A filtered version of the projector is obtained by estimating the P vectors from the K eigenvectors associated with the K highest eigenvalues, with K<P.
- setting to zero probability values in the projector that are below a given threshold [10]: threshold values were expressed as a percentage of the maximum probability value in the projector, and were varied from 0.1% and 10%.

Two types of errors in modeling the projector **R** were considered for the cubic phantom. First, a 4 keV shift in the energy calibration of the camera was mimicked by reconstructing data recorded in the 122-150 keV energy range using a projector calculated for the 126-154 keV energy range. Second, a shift in spatial sampling was introduced by introducing a 2 mm or 5 mm off-set horizontally and vertically between the grid used to calculate the projections and that used to calculate the projector.

### III. RESULTS AND DISCUSSION

*A. Effect of noise in the projector and filtering procedures*

Table 1 shows how increasing the number of simulated counts used to estimate the projector improves spatial resolution and signal-to-noise ratio, without affecting the quantitative results much. PCA filtering also improves spatial resolution and signal-to-noise ratio, together with relative quantitation, suggesting that high statistics simulations combined with PCA filtering should be used to achieve the best results. Setting low probability values in the projector to zero did not appear as an efficient way for reducing noise in the projector.

Results on the cubic phantom suggested that, when filtering the projector, $r_{ij}$ values should be estimated with a relative uncertainty at least inferior to 0.003 % to yield excellent results with F3DMC. Using this empirical rule for the cylinder phantom, one can roughly estimate the minimum number of photons that should be simulated to get a robust enough projector. Assuming a uniform distribution of detected counts, about 1300 billion counts should be simulated to get a robust projector, i.e. a projector with a statistics about 20 times higher than the one considered in this study.

### B. Robustness of F3DMC with respect to modeling error

Table 2 shows that F3DMC results were almost not affected by what was supposed to mimic a 4 keV shift in the energy calibration of the gamma camera ($P_1$). They were weakly affected by the 2 mm off-set between the sampling grid used to calculate the projector and the sampling grid used to calculate the projection ($P_2$). The results strongly deteriorate for a 5 mm shift ($P_3$), consistent with the fact that the sources in the phantom were 1 cm thick.

### C. Cylindrical phantom

Ideal images and images reconstructed using the F3DMC are shown in Figure 2, each image being scaled so that its maximum value corresponds to the maximum value of the color scale. Quantitative indices are given in Table 3.

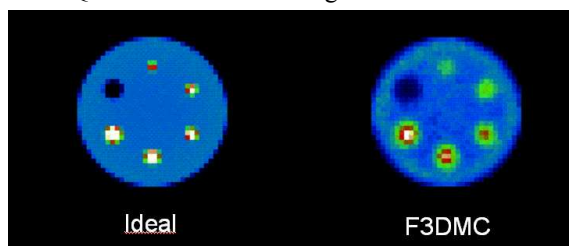

Fig. 2. Slice through the original cylindrical phantom and corresponding slice reconstructed with F3DMC.

Results show that image sampling itself decreases activity ratios compared to their ideal value, from 55% for the smallest hot cylinder to 33% for the biggest hot cylinder. Considering the value measured on the spatially sampled activity maps as a reference, F3DMC yielded accurate activity ratio estimates for sphere sizes down to 9.5 mm in diameter, suggesting a local spatial resolution in the reconstructed image of about 5 mm.

## IV. CONCLUSION

Using the cubic phantom, we have shown the importance of calculating the projector from a high statistics simulation and the added value of reducing noise in the projector by appropriate filtering. In addition, F3DMC appears robust enough with respect to small errors in energy calibration of the gamma camera and small shifts between the voxel sampling used to calculate the projector and the sampling of the projections. The feasibility of reconstructing 64 projections 64x64 of a realistic object using F3DMC has also been proven with encouraging results, in terms of spatial resolution recovery and quantification.

TABLE 1
EFFECT OF NOISE AND FILTERING BY PCA OR THRESHOLD ON QUANTITATIVE CRITERIA CHARACTERIZING THE CUBIC PHANTOM IMAGES (VALUES GIVEN IN PARENTHESES CORRESPOND TO RELATIVE ERRORS BETWEEN IDEAL AND ESTIMATED ACTIVITY RATIOS)

| Reconstruction method | Spatial resolution (in plane / axial) | Activity ratios | | Absolute activity concentration (%RA) | SNR |
| --- | --- | --- | --- | --- | --- |
| | | Z:Y | Z:X | | |
| Ideal | 1.00 / 1.00 | 1.33 | 2.00 | 100 | - |
| F3DMC 1 billion | 2.09 / 1.25 | 1.21 (-9.1%) | 1.70 (-15%) | 100.5 | 167 |
| F3DMC 30 billion | 1.08 / 1.02 | 1.43 (7.6%) | 1.63 (-18.5%) | 103.1 | 210 |
| F3DMC 1 billion + PCA filtering | 1.22 / 1.09 | 1.37 (3.0%) | 1.70 (-15%) | 85.3 | 175 |
| F3DMC 30 billion + PCA filtering | 1.03 / 1.01 | 1.33 (0%) | 1.77 (-11.5%) | 92.6 | 247 |
| F3DMC 30 billion + values < 0.1% max set to 0 | 1.08 / 1.02 | 1.43 (7.6%) | 1.62 (-19.0%) | 103 | 210 |
| F3DMC 30 billion + values < 1% max set to 0 | 1.09 / 1.03 | 1.43 (7.6%) | 1.61 (-19.5%) | 108.5 | 211 |
| F3DMC 30 billion + values < 10% max set to 0 | 1.13 / 1.14 | 1.50 (12.8%) | 1.50 (-25.0%) | 150 | 197 |

TABLE 2
QUANTITATIVE CRITERIA CHARACTERIZING THE CUBIC PHANTOM IMAGES WITH RESPECT TO MODELING ERRORS (VALUES GIVEN IN PARENTHESES CORRESPOND TO RELATIVE ERRORS BETWEEN IDEAL AND ESTIMATED ACTIVITY RATIOS)

| Reconstruction method | Spatial resolution (in plane / axial) | Relative quantitation | | Absolute quantitation (%RA) | SNR |
|---|---|---|---|---|---|
| | | Z:Y | Z:X | | |
| Ideal | 1.00 / 1.00 | 1.33 | 2.00 | 100 | - |
| F3DMC (P) | 1.03 / 1.01 | 1.33 (0%) | 1.77 (-11.5%) | 92.6 | 247 |
| F3DMC ($P_1$) | 1.03 / 1.01 | 1.32 (-0.8%) | 1.82 (-9.9%) | 92.8 | 257 |
| F3DMC ($P_2$) | 1.06 / 1.07 | 1.46 (9.8%) | 1.76 (-12%) | 89.9 | 181 |
| F3DMC ($P_3$) | 1.43 / 2.00 | 2.15 (61.7%) | 1.65 (-17.5%) | 86.7 | 102 |

TABLE 3
QUANTITATIVE CRITERIA CHARACTERIZING THE CYLINDRICAL PHANTOM IMAGES RECONSTRUCTED WITH F3DMC (VALUES GIVEN IN PARENTHESES CORRESPOND TO RELATIVE ERRORS BETWEEN ESTIMATED ACTIVITY RATIOS AND ACTIVITY RATIOS OBTAINED AFTER SAMPLING)

| Reconstruction method | Relative quantitation | | | | | | Absolute quantitation |
|---|---|---|---|---|---|---|---|
| | Cyl 1 4.8 mm | Cyl 2 6.4 mm | Cyl 3 7.8 mm | Cyl 4 9.5 mm | Cyl 5 11.1 mm | Cyl 6 12.7 mm | |
| Ideal | 4.00 | 4.00 | 4.00 | 4.00 | 4.00 | 0.00 | 100% |
| After sampling | 1.82 | 2.16 | 2.39 | 2.47 | 2.65 | 0.39 | 100% |
| F3DMC | 1.16 | 1.43 | 1.65 | 1.84 | 2.04 | 0.52 | 98.9% |
| F3DMC + PCA filtering | 1.38 (-24.2%) | 1.73 (-19.9%) | 2.05 (-14.3%) | 2.35 (-4.9%) | 2.66 (0.4%) | 0.38 (-2.6%) | 79.9% |